\documentclass[journal]{IEEEtran}

  \usepackage[pdftex]{graphicx}
\usepackage{comment}
\usepackage[font=small,hang]{caption}
\usepackage{amsmath,amssymb} 
\usepackage{color,soul}
\usepackage{float}
\usepackage{booktabs}
\usepackage{multirow}
	\usepackage{textcomp}
\usepackage{array}
\usepackage[comma,sort&compress,numbers]{natbib}
\usepackage[font=small]{caption}
\newcolumntype{P}[1]{>{\centering\arraybackslash}p{#1}}
\newcolumntype{M}[1]{>{\centering\arraybackslash}m{#1}}
\usepackage{url}
\usepackage{siunitx}
\usepackage{scrextend}
   \graphicspath{{./figures/}}
\usepackage[pdftex]{hyperref}

\begin{document}

\title{MFRNet: A New CNN Architecture for Post-Processing and In-loop Filtering}

\author{Di Ma, Fan Zhang,~\IEEEmembership{Member,~IEEE,}
        and David R. Bull,~\IEEEmembership{Fellow,~IEEE}
\thanks{D. Ma, F. Zhang and D. R. Bull are with the Department of Electrical and Electronic Engineering, University of Bristol, Bristol, BS8 1UB, UK (e-mail: di.ma@bristol.ac.uk; fan.zhang@bristol.ac.uk; dave.bull@bristol.ac.uk).}

\thanks{Manuscript submitted June 30, 2020.}}


\maketitle
\begin{abstract}
In this paper, we propose a novel convolutional neural network (CNN) architecture, MFRNet, for post-processing (PP) and in-loop filtering (ILF) in the context of video compression. This network consists of four Multi-level Feature review Residual dense Blocks (MFRBs), which are connected using a cascading structure. Each MFRB extracts features from multiple convolutional layers using dense connections and a multi-level residual learning structure. In order to further improve information flow between these blocks, each of them also reuses high dimensional features from the previous MFRB. This network has been integrated into PP and ILF coding modules for both HEVC (HM 16.20) and VVC (VTM 7.0), and fully evaluated under the JVET Common Test Conditions using the Random Access configuration. The experimental results show significant and consistent coding gains over both anchor codecs (HEVC HM and VVC VTM) and also over other existing CNN-based PP/ILF approaches based on Bj{\o}ntegaard Delta measurements using both PSNR and VMAF for quality assessment. When MFRNet is integrated into HM 16.20, gains up to 16.0\% (BD-rate VMAF) are demonstrated for ILF, and up to 21.0\% (BD-rate VMAF) for PP. The respective gains for VTM 7.0 are up to 5.1\% for ILF and up to 7.1\% for PP.
\end{abstract}


\IEEEpeerreviewmaketitle

\section{Introduction}
\label{sec:intro}

\IEEEPARstart{I}{n} recent years the consumption of video content has increased dramatically. This has been  associated with demands for improved viewing quality, and for more immersive experiences (including augmented and virtual reality (AR and VR)) with multiple views, higher spatial and temporal resolutions and wider dynamic range,  \cite{bull2014communicating}. These all create pressure on network capacity and present significant challenges for video compression. 

To address this, new video coding standards have been initiated including  Versatile Video Coding (VVC) \cite{bross2018versatile}, Alliance for Open Media Video 1 and 2 (AV1/AV2) \cite{av1} and Essential Video Coding (EVC) \cite{choi2020overview}. All of these are expected to achieve significant coding gains compared to current standards \cite{zhang2020comparing,topiwala2019performance} such as High Efficiency Video Coding (HEVC) \cite{hevc}. These new coding standards all employ a similar framework to that used in previous codecs such as H.264/AVC (Advanced Video Coding) \cite{avc}, but with much more sophisticated modifications and enhancements. None of them however, exploit recent advances in artificial intelligence and machine learning.

The last decade has seen significant advances in the application of machine learning, especially using convolutional neural networks (CNNs), for image and video analysis, recognition and processing \cite{yao2019review}. More recently, deep learning techniques have also been applied to the problem of image and video compression, both to enhance existing coding modules and also to provide new end-to-end solutions \cite{ma2019image,liu2020deep}. Amongst approaches that target specific coding modules, CNN-based post-processing (PP) and in-loop filtering (ILF) have have delivered reductions in visual artefacts and overall improvements in perceptual quality, showing bit rate savings against codecs such as VVC. The coding gains reported are primarily for intra coding \cite{dai2017convolutional,wang2019progressive}, and the network architectures employed in many cases do not represent the latest deep learning advances.

In the above context, a new CNN architecture (MFRNet) is proposed for video compression, to enhance both post-processing and in-loop filtering. This employs Multi-level Feature review Residual dense Blocks (MFRB) in a cascading structure. The network was trained on a large database with diverse video content, and integrated into both HEVC (HM 16.20) and VVC (VTM 7.0) reference codecs. Results were evaluated on test sequences from the Joint Video Exploration Team (JVET) Common Test Conditions (CTC) \cite{jvetctc} dataset using the Random Access configuration (the most effective default coding mode in HEVC and VVC). Significant improvements were observed over both HEVC and VVC test models and over other state-of-the-art CNN-based PP and ILF approaches. The proposed architecture also demonstrates superior performance when compared to other popular network structures.

The contributions of this paper are summarised below:

\begin{enumerate}
    \item A novel CNN architecture is presented exploiting multi-level feature review residual dense blocks for post-processing and in-loop filtering.
    \item This network has been integrated into both HEVC and VVC test models, demonstrating significant coding gains for Random Access configurations.
    \item A comprehensive comparison is made between the proposed architecture and thirteen existing popular network structures, based on the same training and evaluation material.
\end{enumerate} 

The remainder of the paper is organised as follows. Section \ref{sec:background} reviews recent advances in video coding standards and the state of the art in deep video compression, in particular for post-processing and in-loop filtering. Section \ref{sec:algorithm} describes the CNN-based PP/ILF coding modules, the proposed network architecture, and the training/evaluation processes, while Section \ref{sec:results} reports experimental results with analysis and discussion. Finally, conclusions and future work are outlined in Section \ref{sec:conclusion}. 

\section{Background}
\label{sec:background}

This section first overviews the latest standardisation activity for video compression (Section \ref{subsec:standards}), and reviews current applications of deep learning in the context of video compression (Section \ref{subsec:dvc}). Section \ref{subsec:ppilf} describes the primary functions of post-processing (PP) and in-loop filtering (ILF), while Section \ref{subsec:cnnppilf} provides a brief summary of existing CNN-based PP and ILF approaches.

\subsection{Video coding standards}
\label{subsec:standards}

Since the early 1980s, multiple generations of video coding standards have been developed by ITU-T and/or ISO/IEC for various application scenarios. Among these, the most successful has been H.264/AVC (Advanced Video Coding) \cite{avc} which was released in 2004 targeting Internet streaming and HDTV. H.264/AVC is still widely used, despite its successor HEVC/H.265 (High Efficiency Video Coding, 2013) \cite{hevc} providing nearly 50\% coding gain. In 2018, in order to provide improved support for immersive video formats (e.g. high dynamic range and 360$^{\circ}$) with further compression efficiency improvements, a new coding standard, Versatile Video Coding (VVC) \cite{bross2018versatile}, was initiated. This is expected to be finalised in 2020 and currently \cite{zhang2020comparing} shows more than 35\% overall coding gain over HEVC, but with a significant increase in encoder complexity.

In parallel with developments under ITU-T and ISO/IEC, the Alliance for Open Media (AOM, an industry consortium) has developed an open source and royalty-free coding solution, AOM Video 1 (AV1). AV1 was launched in 2018 to target Internet streaming and its most recent versions appear to offer evident and consistent performance improvements over HEVC \cite{zhang2020comparing, chen2020overview}. The development of its successor, AV2 (AOMedia Video 2) is also planned to start in 2020. Other recent advances in coding standards include the Essential Video Coding/MPEG-5 \cite{choi2020overview} and AVS standards \cite{avs} that target royalty free solutions and complexity-performance trade-offs.

\subsection{Deep video compression}
\label{subsec:dvc}
Inspired by recent advances in artificial intelligence, deep neural networks have started to play an important role in video compression to enhance individual coding tools including intra coding \cite{yeh2018hevc,li2018fully}, inter prediction \cite{zhao2018enhanced,zhao2019enhanced}, transforms \cite{puri2017cnn,jimbo2018deep}, quantisation \cite{alam2015perceptual}, entropy coding \cite{song2017neural,ma2018convolutional}, formats adaptation \cite{lin2018convolutional,afonso2019video,ma2020gan}, post-processing (PP) \cite{lin2019partition,zhang2020pp} and in-loop filtering (ILF) \cite{zhang2018residualInLoop,jia2019content}. Alternative deep learning coding architectures have also been proposed based on an end-to-end training and optimisation process \cite{balle2016end,rippel2019learned,lu2019dvc,djelouah2019neural,habibian2019video}. More details on deep video compression can be found in this special issue and in \cite{ma2019image,liu2020deep,zhang2020machine}.

For learning-based compression, in order to achieve good model generalisation and avoid potential over-fitting problems, it is essential for training material to include diverse content covering different formats and texture types. Existing learning-based coding methods often employ training databases developed for super-resolution \cite{Agustsson_2017_CVPR_Workshops}, frame interpolation \cite{soomro2012ucf101} or classification \cite{russakovsky2015imagenet}. More recently, a new extensive and representative video database, BVI-DVC, has been made available for training CNN-based coding tools \cite{ma2020bvi}. This database has been shown to provide significant improvements in terms of coding gains over other commonly used training databases for various CNN architectures and coding modules (including PP and ILF).

\subsection{PP and ILF in video coding}
\label{subsec:ppilf}
Following compression, decoded videos often exhibit noticeable visual artefacts including blocking discontinuities, ringing and blurring, and these become more evident as quantisation level increases. To reduce these distortions, post-processing can be employed at video decoder to improve reconstruction quality. When integrated into the encoding loop, filtered frames with fewer artefacts and improved quality can be used as reference for inter prediction \cite{bull2014communicating}.

In the latest version of VVC, the in-loop filtering module consists of three different stages including deblocking filtering (DBF), sample adaptive offset (SAO) and adaptive loop filtering (ALF). DBF is designed to adaptively suppress artefacts along block boundaries using low-pass smoothing filters according to discontinuity levels \cite{JVET-L0414,JVET-P1001}. SAO is invoked after DBF to make a non-linear adjustment that adds offsets to samples based on a look-up table created at the encoder using histogram analysis of signal amplitudes \cite{JVET-Q0441}. ALF is a new feature in VVC, which minimises distortions between the original and reconstructed blocks using an adaptively trained low-pass filter \cite{JVET-Q0665}. 

\subsection{CNN-based PP and ILF}
\label{subsec:cnnppilf}
Deep neural networks, especially deep CNNs, have made significant contributions to single image super-resolution and image restoration. Common architectural features include (from simple to more complex): (i) Simple concatenated convolutional layers \cite{dong2015image,dong2016accelerating,kim2016accurate}; (ii) the addition of deeper residual blocks \cite{tai2017image,lim2017enhanced,ledig2017photo,zhang2018image,ma2019perceptually}; (iii) including dense connections \cite{zhang2018residual,wang2018esrgan}; (iv) with cascading connections \cite{ahn2018fast,ahn2019photo}; and (v) with feature review structures \cite{wang2018dense,wang2019progressive}. Notable examples of the use of CNNs for post-processing and in-loop filtering include  \cite{dai2017convolutional,wang2018dense,ma2019residualpp,zhao2019cnn,lin2019partition,zhang2018residualInLoop,jia2019content,wang2019progressive,zhang2019vistra2,JVET-J0031,JVET-O0063,JVET-O0079,JVET-O0131,JVET-O0132,zhang2020pp,JVET-O0056,JVET-O0101,wang2019attention}. However, most of these methods do not employ advanced features such as residual dense blocks, cascading connections or feature review structures, and are often trained on relatively small databases \mbox{{\cite{Agustsson_2017_CVPR_Workshops,liu2017robust}}}. This can result in sub-optimal and/or non-generic models, and can only offer consistent coding gains for less effective coding configurations, such as All Intra (AI) and Low Delay (LD) modes in HEVC and VVC. For the case of the more commonly used Random Access (RA) configuration  (which  provides much higher coding efficiency than AI and LD) these existing CNN-based PP and ILF approaches, which employ simple network architectures, demonstrate little or no coding performance improvement.

\section{Proposed Algorithm}
\label{sec:algorithm}

This section presents the new CNN architecture, MFRNet, and describes how this is integrated into CNN-based post-processing (PP) and in-loop filtering (ILF) coding modules. The training and evaluation methodologies are also provided in detail.

\subsection{CNN-based PP and ILF coding modules}

\begin{figure}[ht]
\centering
\includegraphics[width=0.88\linewidth]{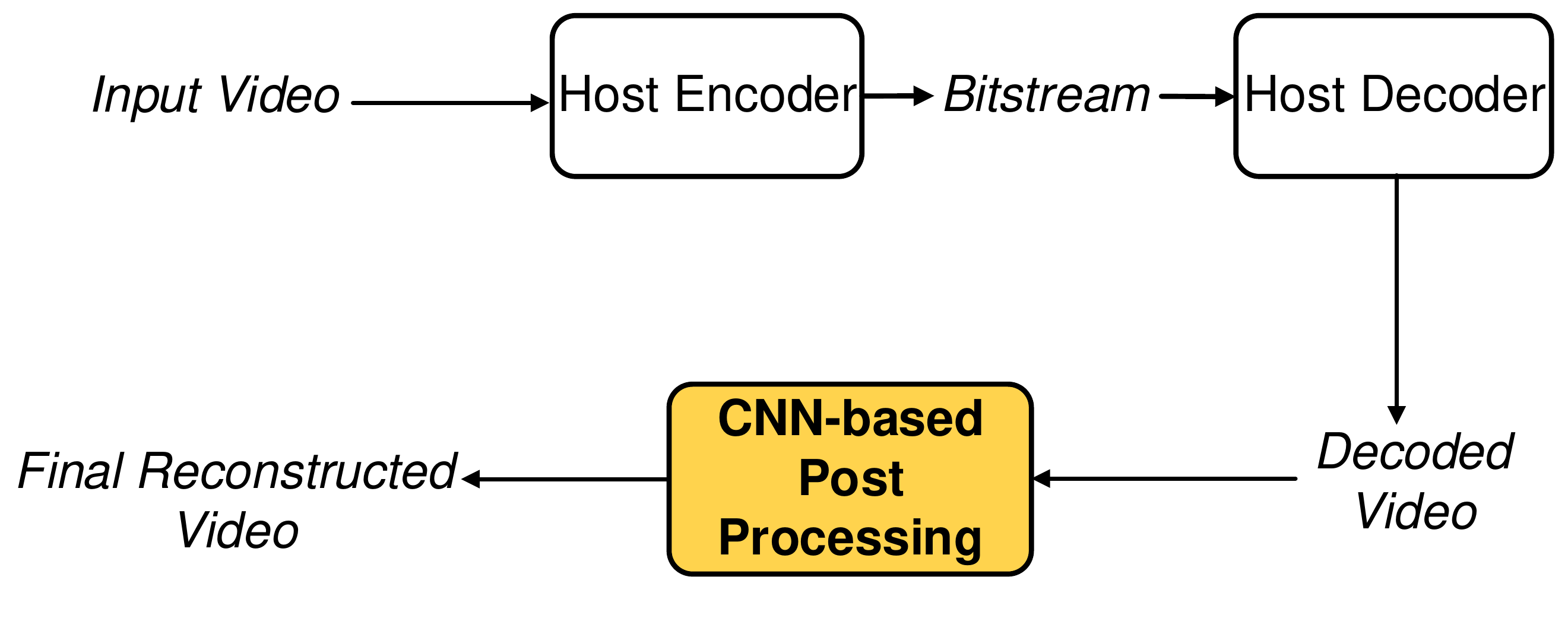}
\caption{Coding workflow with a CNN-based PP module.}
\label{fig:pp}
\end{figure}

\begin{figure}[ht]
\centering
\includegraphics[width=1.01\linewidth]{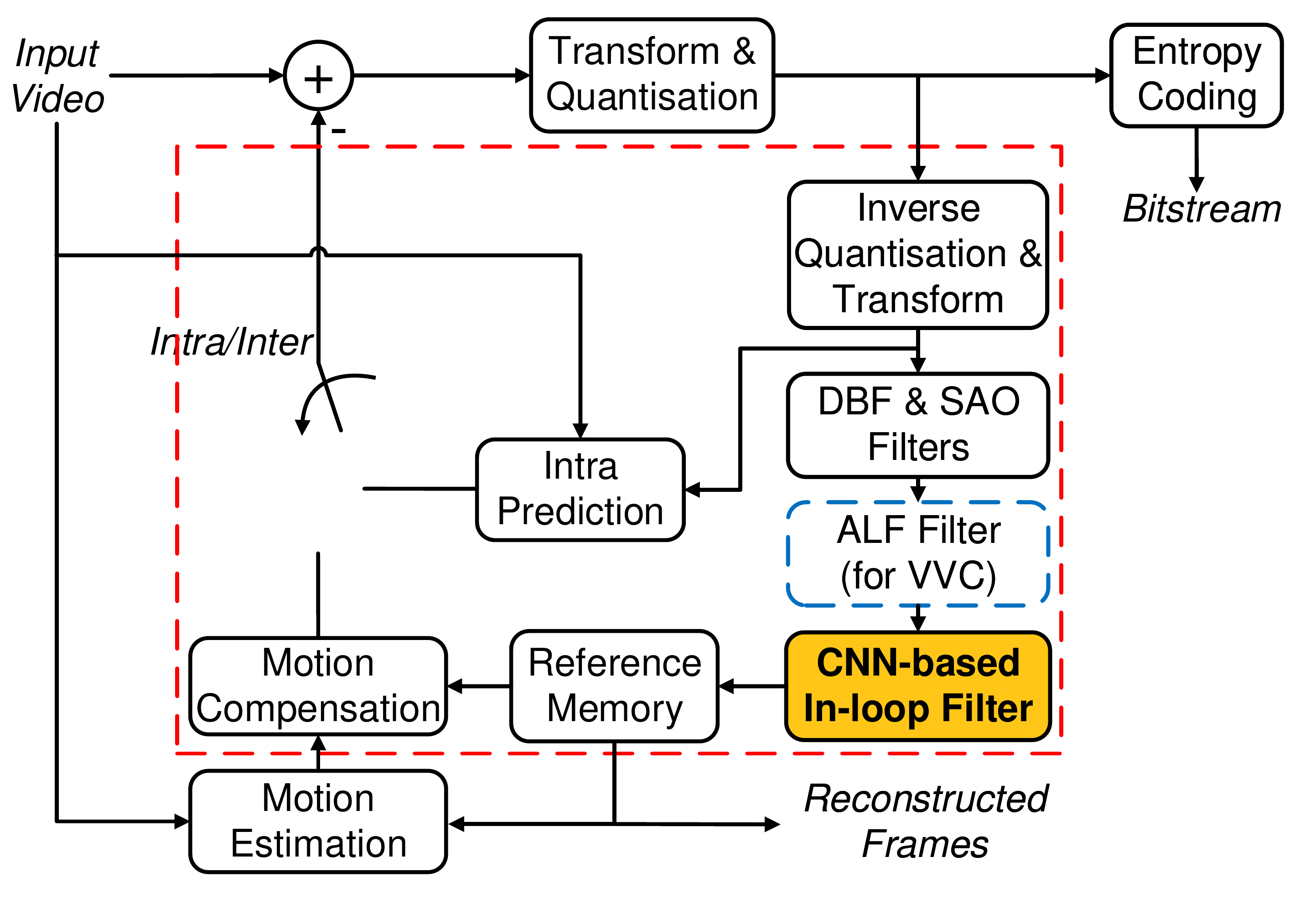}
\caption{Coding workflow with a CNN-based ILF module (the modules in the dashed box form the corresponding decoder).}
\label{fig:ilf}
\end{figure}

The CNN-based PP and ILF modules employed here are illustrated in Fig. \ref{fig:pp} and \ref{fig:ilf}. For the case of post-processing, the employed CNN filter is applied directly after the reconstructed frames are decoded from the transmitted bitstream, and produces a frame with improved reconstruction quality in the same format. As shown in Fig. {\ref{fig:ilf}}, the CNN-based ILF module is located after the conventional in-loop filtering process, and has the same input and output format as for the PP case. This is similar to that for most of the previous contributions on CNN-based ILF filters \mbox{\cite{zhang2018residualInLoop,JVET-O0056}}. Compared to other possible designs, where the CNN operation is not performed as the last step in the whole coding workflow (e.g. before ALF, SAO or DBF), this implementation will not conflict with the existing conventional loop filters and will achieve better reconstruction performance due to its end-to-end optimisation in the training process.

\subsection{Proposed CNN architecture}

\begin{figure*}[htbp]
\centering
\includegraphics[width=1.01\linewidth]{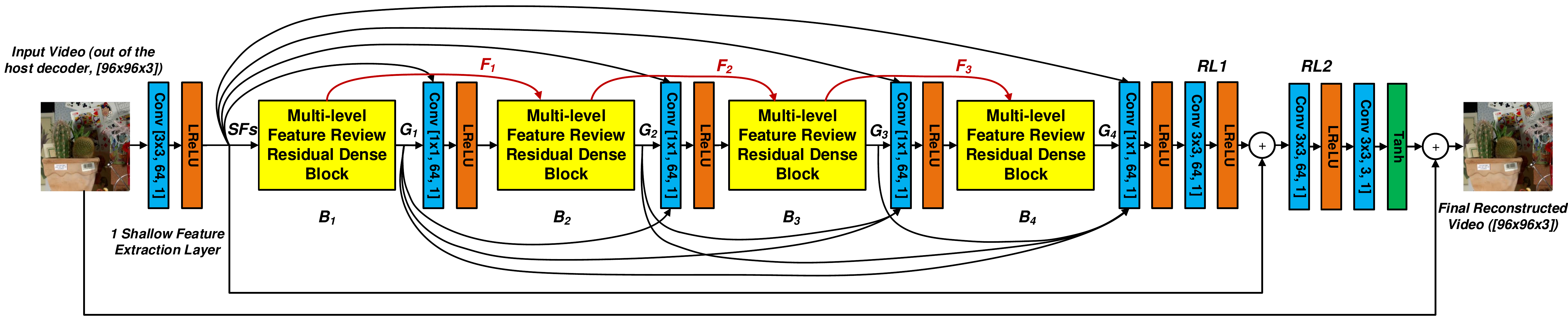}
\caption{Illustration of the proposed MFRNet architecture.}
\label{fig:net}
\end{figure*}

\begin{figure}[htbp]
\centering
\includegraphics[width=1.02\linewidth]{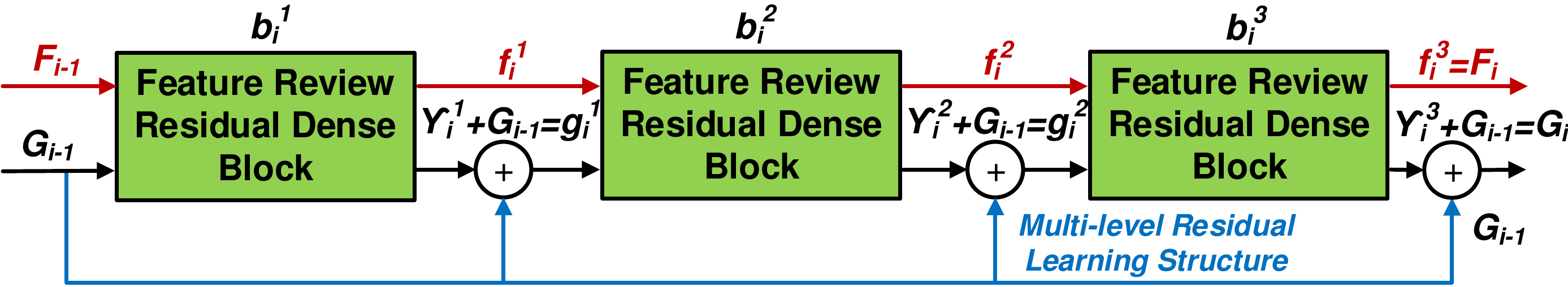}
\caption{Illustration of an MFRB ($B_i$).}
\label{fig:multilevel}
\end{figure}

\begin{figure}[htbp]
\centering
\includegraphics[width=1.02\linewidth]{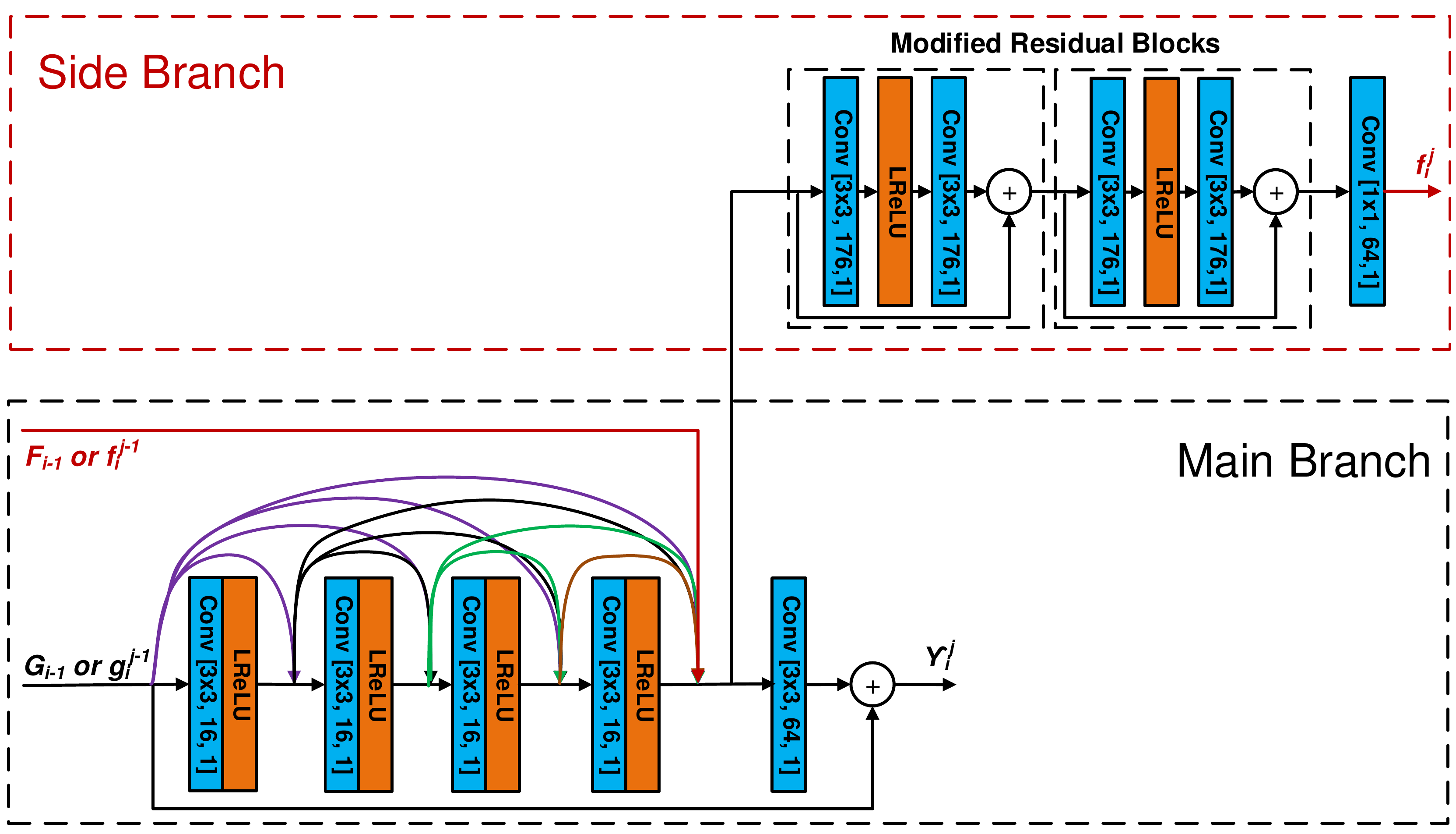}
\caption{Illustration of an FRB ($b_i^j$).}
\label{fig:frrdb}
\end{figure}

The CNN architecture proposed for PP and ILF is illustrated in Fig. \ref{fig:net}-\ref{fig:frrdb}. This network accepts a 96$\times$96 YCbCr 4:4:4 image block as input, and outputs a filtered image block in the same format. It first employs a convolutional layer alongside a Leaky ReLU (LReLU) activation function to extract shallow features (SFs) from the input image block. This SF extraction layer is followed by four Multi-level Feature review Residual dense Blocks (MFRBs, $B1$-$B4$), which are designed for deep dense feature extraction. Ten cascading connections, shown as black curves in Fig. \ref{fig:net}, are utilised to feed the initial SFs and the output from the first three MFRBs ($G_1$, $G_2$ and $G_3$) into following MFRBs or into the first reconstruction layer (shown as RL1 in Fig. \ref{fig:net}) through a 1$\times$1 convolutional layer (with l LReLU). This structure is designed to effectively improve information flow while reducing the number of residual dense blocks in the network \cite{ahn2018fast,ahn2019photo}. Moreover, each of the first three MFRBs also feeds its high dimensional feature outputs, $F_1$, $F_2$, $F_3$, into the next MFRB, as shown in Fig. \ref{fig:multilevel}, in order to reuse previous HDFs \cite{zhang2018residual}. After four MFRBs and the first reconstruction layer (RL1), a skip connection is employed to connect the output of this reconstruction layer and the output of the shallow feature extraction layer. Finally, an additional reconstruction layer (RL2) and an output layer are employed to output a residual signal, which is then combined with the input through a long skip connection to obtain the final image block. The kernel sizes, feature map numbers and stride values for each convolution layer can be found in Fig. \ref{fig:net}. 

Fig. \ref{fig:multilevel} shows the structure of each MFRB ($B_i$, $i$=1, 2, 3 and 4), which contains three Feature review Residual dense Blocks (FRBs), $b_i^1$, $b_i^2$ and $b_i^3$. In many existing CNN architectures, which employ residual (or residual dense) blocks \cite{tai2017image,ledig2017photo,wang2018esrgan,zhang2018residual,ma2019perceptually}, there is only a single information flow, which prevents high level blocks from fully accessing previously generated features. This leads to a problem of diminishing feature reuse, which in turn affects the overall performance of the network \cite{wang2019progressive}. To address this issue, in the proposed architecture, each FRB ($b_i^j$), except the first one in $B_1$ and the last one in $B_4$, is designed to have two inputs and two outputs \cite{wang2019progressive}, as shown in Fig. \ref{fig:multilevel} and \ref{fig:frrdb}. Each FRB not only receives the main branch output from the previous MFRB ($G_{i-1}$) or FRB ($g_i^{j-1}$), but also accepts the side branch output from the previous MFRB ($F_{i-1}$) or FRB ($f_i^{j-1}$), which contains high dimensional features. Respectively, in addition to its main branch output ($\gamma_i^j$), each FRB also feeds its side branch output ($f_i^j$) into the subsequent FRB block in this or the next MFRB (if applicable). This new structure allows each FRB to review the high dimensional features generated in its previous block, which effectively enhances the information flow between blocks. Finally, a multi-level residual learning structure is designed to apply skip connection between the input of the first FRB and the output of each FRB. This enables bypassing of redundant information and stabilises training and evaluation processes \cite{he2016deep,dai2019second}.

 Fig. {\ref{fig:frrdb}} shows the FRB structure ($b_i^j$), which contains a main branch and a side branch. The former first accepts the output from the previous MFRB ($G_{i-1}$) or FRB ($g_i^{j-1}$) if it is available, and extracts dense features through four convolutional layers with dense connections \mbox{\cite{huang2017densely,zhang2018residual}}. Each of these layers contains one convolutional layer and a LReLU function. The output of these four dense convolutional layers are then concatenated together with the side branch output from the previous MFRB ($F_{i-1}$) or FRB ($f_i^{j-1}$) and fed into the last convolutional layer. The output of this layer is combined with the input ($G_{i-1}$ or $g_i^{j-1}$) of this FRB through a skip connection to obtain the final FRB output ($g_i^{j}$). The concatenated, high dimensional features (HDFs) are further fed into two modified residual blocks and one convolutional layer with a 1$\times$1 kernel size to obtain the output of this side branch ($f_i^{j}$) in this FRB. This is also sent to the subsequent FRB block (if applicable) to realise HDF reviewing.

\subsection{Network training and evaluation}

As mentioned in Section \ref{sec:background}, training databases are critical for optimising the performance of learning-based compression algorithms. In this work, a large video database, BVI-DVC \cite{ma2020bvi}, is employed to generate training material. This database contains 800 carefully selected video sequences, all of which have 64 frames with 10 bit and YCbCr 4:2:0 format at four different resolutions from 270p to 2160p. These sequences were compressed by HEVC HM 16.20 and VVC VTM 7.0 codecs using the JVET-CTC Random Access (RA) configuration with four QP values: 22, 27, 32 and 37. During encoding and decoding, all default in-loop filters within the HM and VTM codecs were kept active. For each QP, compressed video frames and their corresponding original counterparts were randomly selected, segmented into 96$\times$96 image blocks, and converted to YCbCr 4:4:4 format. During this process, block rotation was applied to achieve data augmentation. This results in approximately 192,000 pairs of blocks for each QP group.

The proposed CNN was implemented and trained using the  TensorFlow (version 1.8.0) framework with the following training parameters: $\ell 1$ as loss function; Adam optimisation \cite{kingma2014adam} with hyper-parameters of $\beta_1$=0.9 and $\beta_2$=0.999; batch size of 16; 200 training epochs; learning rate  of 0.0001; weight decay of 0.1 for every 100 epochs.

Based on the generated training content, four CNN models (aligned with the four QP groups) were trained for each codec (HM 16.20 and VTM 7.0), which are subsequently used in the evaluation stage for different base QP values:
\begin{equation}
\label{eq1}
{\rm CNN\ Models}=\left\{
\begin{aligned}
{\rm Model_1} & , & {\rm QP_{base}}\leq 24.5\\
{\rm Model_2} & , & {\rm 24.5 < QP_{base}} \leq 29.5\\
{\rm Model_3} & , & {\rm 29.5 < QP_{base}} \leq 34.5\\
{\rm Model_4} & , & {\rm QP_{base}} > 34.5\\
\end{aligned}
\right.
\end{equation}

When the trained CNN models are used for post-processing and in-loop filtering, each reconstructed frame is segmented into 96$\times$96 overlapping blocks with an overlap size of 4 pixels, and converted to YCbCr 4:4:4 format. The CNN output image blocks are then converted to the original format and aggregated in the same way to form the final video frame.

 \section{Results and Discussion}
\label{sec:results}

\begin{table*}[ht]
\centering
\caption{Compression results of the MFRNet-based ILF and PP for HM 16.20.}
\begin{tabular}{l || M{1.35cm}|M{1.35cm}| M{1.35cm}|M{1.35cm}|| M{1.35cm} | M{1.35cm} | M{1.35cm}| M{1.35cm}}
\toprule
\multirow{2}{*}{Class-Sequence} &   \multicolumn{4}{c||}{CNN-based In-loop Filtering (\textit{Full Test})} & 
\multicolumn{4}{c}{CNN-based Post-Processing }\\
\cmidrule{2-9}
\centering
& BD-rate (PSNR)&  BD-PSNR& BD-rate (VMAF)& BD-VMAF  &    BD-rate (PSNR)&BD-PSNR&BD-rate (VMAF) &BD-VMAF\\

\midrule \midrule
A1-Campfire&-6.2\% &+0.12dB&-16.5\%&+1.33& -12.4\% &+0.22dB&-19.2\% &+1.61 \\
A1-FoodMarket4&-8.6\%&+0.29dB&-17.1\%&+1.57& -11.1\% &+0.37dB&-20.6\%  &+1.91\\
A1-Tango2&-11.9\% &+0.17dB&-21.6\%&+1.91& -15.1\% &+0.21dB&-25.2\%  &+2.09\\
\midrule
A2-CatRobot1&-12.6\%&+0.23dB&-21.5\%&+1.73& -17.5\% &+0.33dB&-29.0\% &+2.08\\
A2-DaylightRoad2&-17.1\% &+0.21dB&-25.3\%&+1.82& -21.4\% &+0.27dB&-31.8\% &+2.00 \\
A2-ParkRunning3 &-6.8\%&+0.28dB&-11.9\%&+1.32& -8.9\% &+0.37dB&-12.1\% &+1.36 \\
\midrule \textbf{Class A (2160p)}&-10.5\% &+0.22dB& -19.0\%&+1.61&-14.4\%
&+0.30dB& -23.0\%&+1.84\\
\midrule B-BasketballDrive&-12.4\%&+0.30dB&-11.7\%&+0.92& -14.8\% &+0.35dB&-19.5\% &+1.50\\
B-BQTerrace&-16.5\%&+0.23dB& -25.5\%&+0.67&-20.7\% &+0.29dB& -29.9\% &+0.50\\
B-Cactus&-13.2\%&+0.31dB&-16.0\%&+1.34&-14.6\% &+0.34dB& -21.7\% &+1.63\\
B-MarketPlace&-7.3\%&+0.22dB&-14.9\%&+1.64&-9.6\% &+0.30dB& -19.7\%&+2.05 \\
B-RitualDance&-6.0\%&+0.29dB&-13.8\%&+1.57&-10.7\% &+0.53dB& -18.3\% &+2.15\\
\midrule \textbf{Class B (1080p)} &-11.1\%&+0.27dB&-16.4\%
&+1.23& -14.1\% &+0.36dB& -21.8\%&+1.57 \\
 \midrule C-BasketballDrill&-9.4\%&+0.42dB&-8.1\%&+0.90&-14.4\% &+0.65dB&-16.1\% &+1.86\\
C-BQMall&-10.8\%&+0.44dB&-12.9\%&+0.82&-13.6\% &+0.56dB& -20.9\%&+1.39 \\
C-PartyScene&-7.3\%&+0.31dB&-14.7\%&+1.44&-13.6\% &+0.59dB& -19.9\%&+1.91 \\
C-RaceHorses&-7.8\%&+0.30dB&-12.5\%&+1.02&-10.2\% &+0.39dB& -15.8\% &+1.41\\
\midrule \textbf{Class C (480p)} &-8.8\%&+0.37dB&-12.1\%&+1.05&
-13.0\% &+0.55dB& -18.2\% &+1.64\\
 \midrule D-BasketballPass&-8.8\%&+0.45dB&-11.9\% &+1.58&-12.3\% &+0.64dB& -13.9\% &+1.87\\
D-BlowingBubbles&-7.4\%&+0.31dB&-12.5\%&+1.26&-11.6\% &+0.49dB& -17.9\% &+1.79\\
D-BQSquare&-14.9\%&+0.56dB&-23.6\%&+1.05&-24.1\% &+0.92dB& -32.5\% &+1.48\\
D-RaceHorses&-9.0\%&+0.44dB&-13.1\%&+1.32&-10.7\% &+0.53dB& -15.0\% &+1.62\\
\midrule \textbf{Class D (240p)} &-10.0\%&+0.44dB&-15.3\%&+1.30& -14.7\% &+0.65dB& -19.8\% &+1.69\\
 \midrule \midrule \textbf{Overall} &-10.2\%&+0.30dB&-16.0\%&+1.30& -14.1\% &+0.40dB& -21.0\% &+1.70\\\bottomrule
\end{tabular}
\label{tab:hm}
\end{table*}

All of the enhanced codecs have been fully tested under the JVET Common Test Conditions (CTC) using the Random Access configuration (Main10 profile) with four QP values (22, 27, 32 and 37). As discussed in Section I and Section {\ref{sec:background}}, in order to prove the effectiveness of the proposed CNN architecture, all of the experiments were conducted using the Random Access configuration, which offers significantly better coding efficiency compared to the All Intra and Low Delay modes. Nineteen test sequences from JVET CTC SDR (standard dynamic range) classes A1, A2, B, C and D were used as test material. None of these sequences are included in the training database. The original HM 16.20 and VTM 7.0 codecs are employed as benchmark anchors for all tested approaches, and rate quality performance is measured using the Bj{\o}ntegaard Delta  \cite{BD} methods based on two quality metrics, Peak Signal-to-Noise-Ratio (PSNR, luminance channel only) and Video Multimethod Assessment Fusion (VMAF, version 0.6.1) \cite{li2016toward}. PSNR is widely used as a quality metric for image and video compression while VMAF is a learning-based assessment method, which combines multiple quality metrics and video features using a Support Vector Machine (SVM) regressor. The latter has been shown to offer better correlation performance with subjective opinions on compressed content \cite{zhang2018bvi}. It is also noted that during the evaluation stage, the original in-loop filters in both HM and VTM remain enabled throughout the coding process for all test cases. The training and evaluation processes were both executed on a shared cluster, BlueCrystal Phase 4 (BC4) based in the University of Bristol \cite{bc4}, in which each node contains two 14 core 2.4 GHz Intel E5-2680 V4 (Broadwell) CPUs, 128 GB of RAM, and NVIDIA P100 GPU devices.

\subsection{Compression performance}

\begin{table*}[ht]
\centering
\caption{Compression results of the MFRNet-based ILF and PP for VTM 7.0.}
\begin{tabular}{l || M{1.35cm}|M{1.35cm}| M{1.35cm}|M{1.35cm}|| M{1.35cm} | M{1.35cm} | M{1.35cm}| M{1.35cm}}
\toprule
\multirow{2}{*}{Class-Sequence} &  \multicolumn{4}{c||}{CNN-based In-loop Filtering (\textit{Full Test})} & \multicolumn{4}{c}{CNN-based Post-Processing}\\
\cmidrule{2-9}
\centering
& BD-rate (PSNR) & BD-PSNR&  BD-rate (VMAF) &BD-VMAF&BD-rate (PSNR) &BD-PSNR&   BD-rate (VMAF)&BD-VMAF\\
\midrule \midrule
A1-Campfire&-4.5\%&+0.07dB &-8.3\%&+0.52&  -6.8\%&+0.10dB& -10.3\%&+0.67\\
A1-FoodMarket4& -3.5\% &+0.10dB&-7.3\% &+0.49&-5.0\% &+0.14dB& -9.0\%&+0.68\\
A1-Tango2& -5.9\% &+0.07dB&-6.2\% &+0.40&-8.4\% &+0.09dB& -7.1\%&+0.52\\
\midrule
A2-CatRobot1& -6.7\%&+0.10dB &-6.5\%&+0.36&-9.3\%&+0.13dB& -9.0\%&+0.51\\
A2-DaylightRoad2& -7.6\%&+0.07dB &-8.3\%&+0.32 & -9.5\%&+0.09dB& -10.0\%&+0.32\\
A2-ParkRunning3 & -1.5\% &+0.06dB&-3.1\% &+0.27&-2.7\%&+0.11dB& -3.5\%&+0.38\\
\midrule \textbf{Class A (2160p)}&-5.0\%&+0.08dB 
& -6.6\% &+0.39&-7.0\%&+0.11dB&-8.2\%&+0.51\\
\midrule B-BasketballDrive& -4.3\% &+0.09dB&-5.5\% &+0.31& -7.2\%&+0.15dB& -6.2\%&+0.44\\
B-BQTerrace&-6.9\% &+0.09dB& -5.2\% &+0.10&-8.1\% &+0.10dB& -7.2\%&+0.33\\
B-Cactus&-4.4\% &+0.09dB& -5.4\% &+0.34& -6.7\%&+0.13dB&-7.5\%&+0.51\\
B-MarketPlace&-3.3\%&+0.09dB & -4.4\% &+0.35&-4.4\%&+0.12dB&-5.9\%&+0.51\\
B-RitualDance&-2.8\% &+0.13dB& -3.9\%&+0.38&-5.2\%&+0.24dB& -6.3\%&+0.66\\
\midrule \textbf{Class B (1080p)} & -4.3\% &+0.10dB& -4.9\%&+0.30 &-6.3\%&+0.15dB&-6.6\%&+0.49\\
 \midrule C-BasketballDrill&-4.4\%&+0.18dB &-1.7\% &+0.20& -6.7\%&+0.28dB&-4.8\%&+0.53\\
C-BQMall&-4.2\% &+0.15dB& -5.5\% &+0.28&-7.4\%&+0.27dB&-7.8\%&+0.45\\
C-PartyScene&-2.0\%&+0.08dB & -3.4\% &+0.19&-6.1\%&+0.25dB&-5.4\%&+0.40\\
C-RaceHorses&-2.5\% &+0.09dB& -5.4\% &+0.37&-3.7\%&+0.13dB&-5.7\%&+0.47\\
\midrule \textbf{Class C (480p)} &
-3.3\% &+0.13dB& -4.0\%&+0.26&-6.0\%&+0.23dB&-5.9\%&+0.46\\
 \midrule D-BasketballPass&-6.2\% &+0.30dB& -4.6\% &+0.48&-7.4\%&+0.37dB&-5.5\%&+0.67\\
D-BlowingBubbles&-4.6\% &+0.19dB& -3.6\% &+0.25&-5.9\%&+0.24dB&-5.2\%&+0.43\\
D-BQSquare&-6.6\% &+0.24dB& -3.1\%&+0.12 &-11.5\%&+0.41dB&-12.4\%&+0.25\\
D-RaceHorses&-4.6\%&+0.21dB& -5.6\% &+0.47&-5.7\%&+0.27dB&-5.9\%&+0.58\\
\midrule \textbf{Class D (240p)} & -5.5\% &+0.24dB& -4.2\%&+0.33&-7.6\%&+0.32dB&-7.3\%&+0.48\\
 \midrule \midrule \textbf{Overall} & -4.6\% &+0.10dB& -5.1\% &+0.30&-6.7\%&+0.20dB&-7.1\%&+0.50\\\bottomrule
\end{tabular}
\label{tab:vtm}
\end{table*}

TABLE \ref{tab:hm} and \ref{tab:vtm} summarise the compression performance of the PP and ILF coding modules (with the proposed CNN) when integrated into HEVC HM 16.20 and VVC VTM 7.0. It can be observed that our proposed approach achieves significant and consistent coding gains on all test sequences when integrated into HEVC, with average BD-rates of -10.2\% and -14.1\% for PP and ILF respectively. The coding gains are reduced for VTM, but are still significant with average BD-rates of -4.6\% and -6.7\% for ILF and PP respectively based on the assessment of PSNR. It can also be seen that, for both host codecs and both tested coding modules, the bitrate savings according to VMAF are generally higher than those for PSNR.

As shown in TABLE \ref{tab:hm} and \ref{tab:vtm}, the coding gains for PP are consistently higher than those for ILF, by approximately 2\% for VTM and 4-5\% for HM (in terms of BD-rate). This may at first appear surprising but it should be remembered that, unlike conventional post processing, CNN-based PP does employ end-to-end training. In addition, when CNN-processed frames are employed as a reference (after in-loop filtering), they are used to predict subsequently encoded frames through motion estimation and compensation. This process has not been reflected in the current CNN training (i.e. with CNN-processed content as network input), and is likely to  cause the CNN-based filter to become less effective. Similar results have been observed by other authors when the same CNN is employed for both PP and ILF \cite{JVET-O0063}\footnote{It is also noted that, in TABLE \ref{tab:hmcompare}, the ILF results are better than PP for \cite{wang2018dense,dai2017convolutional}. This is because these CNN models employed for PP have been re-trained using data that is different \cite{lin2019partition} to that in their original literature.}.

\subsection{Comparison between CNN-based PP and ILF approaches}

The coding performance of the proposed CNN model is compared here with other notable CNN-based PP and ILF methods developed for the HEVC and VVC Random Access configuration. These include \cite{dai2017convolutional,ma2019residualpp,zhao2019cnn,wang2018dense,lin2019partition,wang2018dense,zhang2016low,zhang2018residualInLoop,jia2019content,wang2019attention,JVET-O0063,JVET-O0079,JVET-O0131,JVET-O0132}\footnote{It is noted that \cite{zhang2016low} has been commonly used as a benchmark for ILF approaches, although it is not a CNN-based solution. We have included it here due to its consistent performance and popularity.}. It should be noted that these approaches have not been re-implemented due primarily to a lack of their source code. Instead their compression results are extracted directly from the corresponding literature..

TABLE \ref{tab:hmcompare} and \ref{tab:vtmcompare} summarise BD-rate (PSNR) results for five PP and five ILF methods (described above) for each host codec (HM and VTM) and compare with our approach. Due to the limitations of results available in the literature, only results for Class C and D are compared for HEVC HM. It can be observed that, for both host codecs and for the two coding modules, when MFRNet is integrated into PP and ILF modules, it significantly outperforms competing methods, and the improvements are consistent across content classes. This is likely due to the advanced structures employed in the proposed MFRNet architecture and the diversity of the training content used.

\begin{table*}[htbp]
\centering
\caption{Comparison between MFRNet-based PP and ILF and existing CNN-based PP and ILF approaches for HEVC.}
\begin{tabular}{l || M{1.85cm} || M{2.15cm}|| M{2.15cm}|| M{2.15cm} || M{1.85cm} || M{2.15cm} }
\toprule
\multicolumn{7}{c}{Compression Performance Comparisons for Post-Processing Tools (HEVC HM)}\\
\midrule
\multirow{2}{*}{Sequence (Class)} &  \cite{dai2017convolutional} (HM 16.9)&\cite{wang2018dense} (HM 16.15))& \cite{ma2019residualpp} (HM 16.0)&\cite{zhao2019cnn} (HM 16.19)& \cite{lin2019partition} (HM 16.0)&{Proposed Method (HM 16.20)}\\
\cmidrule{2-7}
\centering
& BD-rate (PSNR) &   BD-rate (PSNR) &    BD-rate (PSNR)&BD-rate (PSNR)&BD-rate (PSNR) &   BD-rate (PSNR)\\
\midrule
\textbf{Class C (480p)} &0.63\%&-2.6\%&
-6.8\% & -6.6\%&-7.1\%&\textbf{-13.0\%}\\
\midrule \textbf{Class D (240p)} &1.73\%&-2.6\%& -8.0\% & -4.8\%&-7.3\%&\textbf{-14.7\%}\\
 \midrule \textbf{Overall} &1.2\%&-2.6\%& -7.4\% & -5.7\% &-7.2\%&\textbf{-13.9\%}\\
\midrule
\multicolumn{7}{c}{Compression Performance Comparisons for In-loop Filtering Tools (HEVC HM)}\\
\midrule
\multirow{2}{*}{Sequence (Class)} &  \cite{zhang2016low} (HM 16.5)&\cite{dai2017convolutional} (HM 16.9)& \cite{wang2018dense} (HM 16.15)& \cite{zhang2018residualInLoop} (HM 12.0)&\cite{jia2019content} (HM 16.9)&{Proposed Method (HM 16.20)}\\
\cmidrule{2-7}
\centering
& BD-rate (PSNR) &   BD-rate (PSNR) &    BD-rate (PSNR)&BD-rate (PSNR)&BD-rate (PSNR) &   BD-rate (PSNR)\\
\midrule \midrule
\textbf{Class C (480p)} &-4.6\%&-3.0\%&
-3.9\% & -7.1\%&-4.5\%&\textbf{-8.8\%}\\
\midrule \textbf{Class D (240p)} &-2.5\%&-2.3\%& -4.6\% & -4.4\%&-3.3\%&\textbf{-10.0\%}\\
  \midrule \textbf{Overall} &-3.6\%&-2.7\%& -4.3\% & -5.8\% &-3.9\%&\textbf{-9.4\%}
 \\\bottomrule
\end{tabular}
\label{tab:hmcompare}
\end{table*}

\begin{table*}[htbp]
\centering
\caption{Comparison between MFRNet-based PP and ILF and existing CNN-based PP and ILF approaches for VVC.}
\begin{tabular}{l || M{1.85cm} || M{2.15cm}|| M{2.15cm}|| M{2.15cm} || M{1.85cm} || M{2.15cm} }
\toprule
\multicolumn{7}{c}{Compression Performance Comparisons for Post-Processing Tools (VVC VTM)}\\
\midrule
\multirow{2}{*}{Sequence (Class)} &  \cite{JVET-O0063} (VTM 5.0)& \cite{JVET-O0079} (VTM 5.0)&\cite{JVET-O0131} (VTM 5.0)&\cite{JVET-O0132} (VTM 5.0)& \cite{zhang2020pp} (VTM 4.0.1)&{Proposed Method} (VTM 7.0)\\
\cmidrule{2-7}
\centering
& BD-rate (PSNR) &   BD-rate (PSNR) &    BD-rate (PSNR)&BD-rate (PSNR)&BD-rate (PSNR) &BD-rate (PSNR)\\
 \midrule
 \textbf{Class A (2160p)}&-2.0\% & -1.3\%&-1.2\% 
& -0.2\% &-3.3\%&\textbf{-7.0\%}\\
\midrule \textbf{Class B (1080p)} &-1.3\%&-1.5\%
& 0.4\% & -0.2\% &-2.6\%&\textbf{-6.3\%}\\
\midrule \textbf{Class C (480p)} &0.3\%&-3.3\%&
2.2\% & -0.6\%&-3.9\%&\textbf{-6.0\%}\\
\midrule \textbf{Class D (240p)} &N/A&-5.0\%& 6.6\% & -0.8\%&-5.8\%&\textbf{-7.6\%}\\
  \midrule \textbf{Overall} &-1.2\%&-2.6\%& 1.6\% & -0.4\% &-3.8\%&\textbf{-6.7\%}\\
\midrule
\multicolumn{7}{c}{Compression Performance Comparisons for In-loop Filtering Tools (VVC VTM)}\\
\midrule
\multirow{2}{*}{Sequence (Class)} &  \cite{JVET-O0056} (VTM 5.0)&\cite{JVET-O0063} (VTM 5.0)& \cite{JVET-O0079} (VTM 5.0)&\cite{JVET-O0101} (VTM 5.0)&\cite{wang2019attention} (VTM 4.0)& {Proposed Method} (VTM 7.0)\\
\cmidrule{2-7}
\centering
& BD-rate (PSNR) &   BD-rate (PSNR) &    BD-rate (PSNR)&BD-rate (PSNR)&BD-rate (PSNR) &   BD-rate (PSNR)\\ \midrule
\textbf{Class A (2160p)}&-2.0\% & -1.7\%&-0.4\% 
& -1.3\% &N/A&\textbf{-5.0\%}\\
\midrule \textbf{Class B (1080p)} &-1.4\%&-0.6\%
& 0.6\% & -0.8\% &-1.5\%&\textbf{-4.3\%}\\
\midrule \textbf{Class C (480p)} &0.2\%&0.3\%&
-1.2\% & -0.9\%&-3.1\%&\textbf{-3.3\%}\\
\midrule \textbf{Class D (240p)} &N/A&N/A& -3.1\% & -0.8\%&-3.9\%&\textbf{-5.5\%}\\
\midrule \textbf{Overall} &-1.2\%&-0.8\%& -0.9\% & -1.0\% &-2.7\%&\textbf{-4.6\%} 
\\\bottomrule
\end{tabular}
\label{tab:vtmcompare}
\end{table*}

\subsection{Comparisons with other popular CNN architectures}

To further demonstrate the effectiveness of the proposed MFRNet CNN structure, we have also compared it with thirteen popular CNN architectures in the context of PP and ILF for HEVC HM. These include SRCNN \cite{dong2015image}, Highway Networks \mbox{\cite{srivastava2015highway,srivastava2015training}}, FSRCNN \cite{dong2016accelerating}, VDSR \cite{kim2016accurate}, DRRN \cite{tai2017image}, EDSR \cite{lim2017enhanced}, SRResNet \cite{ledig2017photo}, ESRResNet \cite{wang2018esrgan}, CARN {\cite{ahn2018fast}}, RCAN \cite{zhang2018image}, RDN \cite{zhang2018residual}, MSRResNet \cite{ma2019perceptually} and U$^2$-Net {\cite{qin2020u2}}. Most of these models have been widely used in image super-resolution and restoration, and some (VDSR and MSRResNet) have also been utilised in CNN-based video compression tools \cite{afonso2019video,ma2019perceptually,zhang2019BD,zhang2019vistra2}. Most of these approach provided superior performance to the state of the art in their application domain when they were first proposed.

All thirteen models have been re-implemented using the same framework (TensorFlow 1.8.0) and were integrated into PP and ILF coding modules for HEVC HM 16.20.  During re-implementation, the input and output interfaces of these networks have been modified to satisfy the data format requirements. All networks were also trained on the BVI-DVC database following the same methodology as for the proposed network, using loss functions as described in their original literature. Evaluation results on all 19 JVET test sequences are summarised in TABLE \ref{tab:10cnn} and compared to those for  MFRNet. The original HEVC HM 16.20 is employed as a benchmark. It should be noted that a \textit{Short Test} was conducted for evaluating different ILF coding modules as described in JVET proposal M0904 \cite{JVET-M0904}, in which only the first intra period of each test sequence was encoded, while a \textit{Full Test} (processing all frames in the sequence) was applied for PP. The relative computational complexity for each approach has also been calculated and benchmarked against the original HEVC HM 16.20 encoder (for ILF) and decoder (for PP).  

\begin{table*}[ht]

\centering
\caption{Comparison between thirteen popular CNN architectures and the proposed MFRNet in the context of ILF and PP.}
\begin{tabular}{l ||M{1.85cm}|M{1.85cm}|M{1.85cm}||M{1.85cm}|M{1.85cm}|M{1.85cm}}
\toprule
\multirow{2}{*}{CNN Model} & \multicolumn{3}{c||}{CNN-based In-loop Filtering (\textit{Short Test})}&
\multicolumn{3}{c}{CNN-based Post-Processing}\\
\cmidrule{2-7}
\centering
\scriptsize
&   BD-rate  &    BD-rate  & Relative Complexity &BD-rate&    BD-rate & Relative Complexity \\
&  (PSNR)  &  (VMAF)  &(Encoding)&(PSNR)&    (VMAF) & (Decoding)\\
\midrule \midrule
SRCNN \cite{dong2015image}&-1.4\% & -8.5\% & 1.2$\times$ &-1.9\% & -7.4\% &26.5$\times$ \\
FSRCNN \cite{dong2016accelerating}&-1.3\% & -8.1\% & 1.5$\times$ &-1.6\% & -7.3\% &36.2$\times$\\
VDSR \cite{kim2016accurate}&-2.2\% & -6.5\% & 1.8$\times$ &-1.9\% &-7.6\% & 54.3$\times$\\
DRRN \cite{tai2017image}&-6.8\% & -11.0\% & 2.4$\times$ &-10.8\% & -14.9\% &71.6$\times$\\
EDSR \cite{lim2017enhanced}&-5.9\% & -9.9\% & 9.4$\times$ &-10.0\% & -14.6\% &119.3$\times$\\
SRResNet \cite{ledig2017photo}&-6.4\% & -10.6\% & 2.0$\times$ &-9.8\% & -12.7\% &64.9$\times$\\
MSRResNet \cite{ma2019perceptually}&-6.4\% & -11.3\% & 2.1$\times$ &-10.4\% & -14.2\% & 65.1$\times$\\
CARN {\cite{ahn2018fast}}&-6.9\% & -11.1\% & 1.9$\times$ &-11.2\% & -15.4\% & 59.2$\times$\\
U$^2$-Net {\cite{qin2020u2}}&-7.1\% & -11.2\% & 3.3$\times$ &-11.5\% & -15.9\% & 80.2$\times$\\
ESRResNet \cite{wang2018esrgan}&-7.3\% & -12.0\% & 7.8$\times$ &-11.8\% & -17.7\% &101.1$\times$\\
Highway Networks \mbox{\cite{srivastava2015highway,srivastava2015training}}&-7.2\% & -12.7\% & 8.6$\times$ &-12.0\% & -18.1\% & 117.2$\times$\\
RCAN \cite{zhang2018image}&-7.4\% & -11.4\% & 12.4$\times$ &-12.1\% & -18.5\% & 127.6$\times$\\
RDN \cite{zhang2018residual}&-7.5\% & -11.8\% & 5.7$\times$ &-12.2\%& -17.0\% & 91.8$\times$\\
\midrule
MFRNet&\textbf{-9.9\%} & \textbf{-15.6\%} & 5.3$\times$ &\textbf{-14.1\%}& \textbf{-21.0\%} &81.2$\times$
\\\bottomrule
\end{tabular}
\label{tab:10cnn}
	\end{table*}

It can be observed that MFRNet offers the best performance for both PP and ILF when compared to the other thirteen architectures, with average coding gains of 14.1\% for PP and 9.9\% for ILF based on PSNR, and 21.0\% for PP and 15.6\% for ILF according to VMAF. These figures are consistently greater than those for other networks. In contrast, the computational complexity of the proposed architecture is lower than that of Highway Networks, RCAN, EDSR, ESRResNet and RDN.

\subsection{Complexity analysis}

Finally, the complexity figures of the MFRNet-based ILF and PP are presented in TABLE \ref{tab:complexity}. It can be observed that, due to the integration of MFRNet, the encoding complexity (for ILF) is on average 5.3 and 2.4 times comparing to original HEVC HM 16.20 and VVC VTM 7.0 respectively. When the proposed network is employed for post-processing at the decoder, the average decoding time is 81.2 times that of HM and 72.6 times that of VTM.  

\begin{table}[ht]
\centering
\caption{Relative Complexity of the MFRNet-based ILF and PP.}
\begin{tabular}{l|| M{1.11cm}| M{1.11cm}|| M{1.11cm}|M{1.11cm}}

\toprule
\multirow{2}{*}{Hose Codec} & \multicolumn{2}{c||}{HEVC HM 16.20} & \multicolumn{2}{c}{VVC VTM 7.0}\\
\cmidrule{2-5}
\centering
&   ILF &    PP & ILF & PP \\
&   (Encoding) &    (Decoding) &(Encoding)&(Decoding)\\
 \midrule
 \textbf{Class A (2160p)}&1.3$\times$ & 27.3$\times$&1.1$\times$&23.7$\times$\\
\midrule \textbf{Class B (1080p)} &3.9$\times$&51.5$\times$&1.3$\times$&47.2$\times$\\
\midrule \textbf{Class C (480p)} &8.2$\times$&118.7$\times$&2.1$\times$&101.4$\times$\\
\midrule \textbf{Class D (240p)} &10.3$\times$&161.7$\times$&6.1$\times$&148.9$\times$\\
  \midrule \textbf{Average} &5.3$\times$&81.2$\times$&2.4$\times$&72.6$\times$
\\\bottomrule
\end{tabular}
\label{tab:complexity}
\end{table}

\section{Conclusion}
\label{sec:conclusion}

In this paper, a new CNN architecture, MFRNet, has been proposed as a means of enhancing post-processing (PP) and in-loop filtering (ILF) in the context of video compression. MFRNet comprises four multi-level feature review residual dense blocks, and employs a cascading structure to improve information flow. Each of these block is designed to have a dense connection structure to extract features from multiple convolutional layers, and it can also reuse high dimensional features from the previous block. The proposed CNN has been integrated into PP and ILF modules for both HEVC and VVC standard codecs, and has been fully evaluated using the JVET standard test sequences. The results demonstrate significant coding gains, with a 16.0\% improvement for ILF and 21.0\% for PP over HM 16.20 in terms of VMAF, and a corresponding  5.1\% for ILF and 7.1\% for PP against VTM 7.0. Further comparisons have shown the superiority of the MFRNet architecture over other existing popular deep networks, and over other reported CNN-based PP/ILF approaches. Future work will investigate enhanced ILF training strategies and reductions in computational complexity of the proposed network.

\small
\bibliographystyle{IEEEbib}
\bibliography{refs}

\end{document}